# Collaborative Metric Learning Recommendation System: Application to Theatrical Movie Releases

Miguel Campo, JJ Espinoza, Julie Rieger, Abhinav Taliyan
Twentieth Century Fox Film Corporation
{MiguelAngel.Campo-Rembado, JJ.Espinoza, Julie.Rieger, Abhinav.Taliyan}@fox.com

**ABSTRACT**

Product recommendation systems are important for major movie studios during the movie greenlight process and as part of machine learning personalization pipelines. Collaborative Filtering (CF) models have proved to be effective at powering recommender systems for online streaming services with explicit customer feedback data. CF models do not perform well in scenarios in which feedback data is not available, in 'cold start' situations like new product launches, and situations with markedly different customer tiers (e.g., high frequency customers vs. casual customers). Generative natural language models that create useful theme-based representations of an underlying corpus of documents can be used to represent new product descriptions, like new movie plots. When combined with CF, they have shown to increase the performance in 'cold start' situations. Outside of those cases though in which explicit customer feedback is available, recommender engines must rely on binary purchase data, which materially degrades performance. Fortunately, purchase data can be combined with product descriptions to generate meaningful representations of products and customer trajectories in a convenient product space in which proximity represents similarity (in the case of product-to-product comparisons) and affinity (in the case of customer-to-product comparisons). Learning to measure the distance between points in this space can be accomplished with a deep neural network that trains on customer histories and on dense vectorizations of product descriptions. We developed a system based on Collaborative (Deep) Metric Learning (CML) to predict the purchase probabilities of new theatrical releases. We trained and evaluated the model using a large dataset of customer histories spanning multiple years, and tested the model for a set of movies that were released outside of the training window. Initial experiments show gains relative to models that don't train on collaborative preferences.

**CCS Concepts**

Computing Methodologies; Machine Learning; Neural Networks; Information Systems; Recommender Systems

**Keywords**

Collaborative Filtering, Discriminative Deep Metric Learning, Movie Recommendations

## 1. INTRODUCTION

Product recommendation systems are important for movie studios because they help quantify and rank customers' preferences. Fine-grained understanding of customers' preferences is essential early on during the movie greenlight process to simulate possible financial outcomes of plausible narrative options of the same movie script. Closer to the date of the theatrical release, product recommendation systems are becoming a crucial element of studios' personalization and acquisition campaigns.

The accuracy of the model recommendations is important for product launches because acting on erroneous recommendation leads to bad investment decisions, misguided creative messaging and inefficient off-target media and promotion campaigns. Collaborative filtering models (CF) have gained popularity as the engine of recommendation systems for online streaming services [12]. Traditional CF models are not very well equipped though to make recommendations in cold-start situations, e.g., deciding whether to invest in a new movie for a certain audience, and/or recommending the new movie to a given customer. This is especially true for products that are 'novel', like non-sequels movies and movies that cross genres.

Hybrid CF approaches like Collaborative Topic Regression (CTR) and Collaborative Deep Learning (CDL) have been proposed recently to address some of those limitations ([5], [7], [8], [9], [11]). Hybrid models are effective because they use customer histories *and* product content data to jointly determine the latent product vectors and the latent customer vectors. In the inference step, these models show better performance for cold-start recommendations than traditional CF models because they are trained to interpolate content information and collaborative preferences information. If preferences are not available for a new product, they simply lean on the latent representation of the content information.

While the hybrid CF approach works well for products that fall easily into established product categories, it may underperform for products that cross categories. That's in part because (Hybrid) CF models rely on a dot-product operator to determine the level of affinity between customer and products. Dot-product operators do not preserve the triangle inequality [10] and can miss subtle details of the distribution of customers' preferences. In the movie industry, one recent example is the movie The Greatest Showman (TGS). Customers that went to TGS had also gone in the past to Beauty & the Beast (B&B) and Wonder Woman (WW). In a hybrid CF model, the latent vector for TGS is aligned with the latent vector for B&B because of content similarity (music, romance), but not necessarily with WW (dissimilar content). A CF model would successfully recommend TGS to B&B audiences but could fail to make a cold-start recommendation to WW audiences. That's because the content vectors for TGS and WW don't have to be aligned in a trained CF model that has never seen a large movie with a blended B&B-WW audience. Mathematically, the trained CF model doesn't require B&B and WW latent vectors to be aligned because it can learn the preferences of the intersection of those two audiences by simply adjusting the vector 'lengths' of those audiences relative to the other audiences (see further details in [10].)

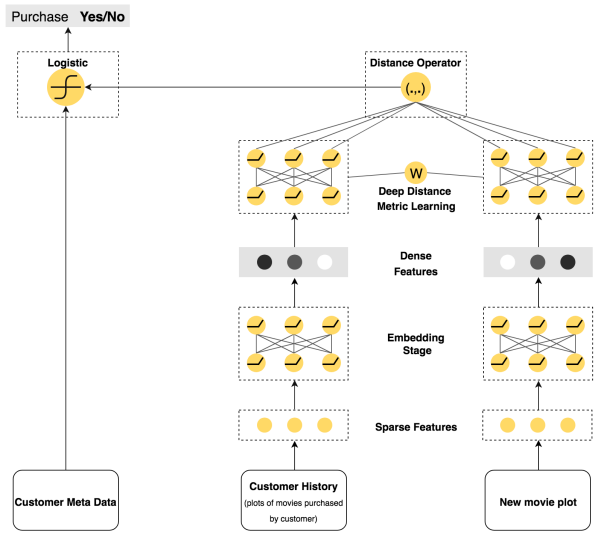

**Figure 1. Architecture of the prediction system**

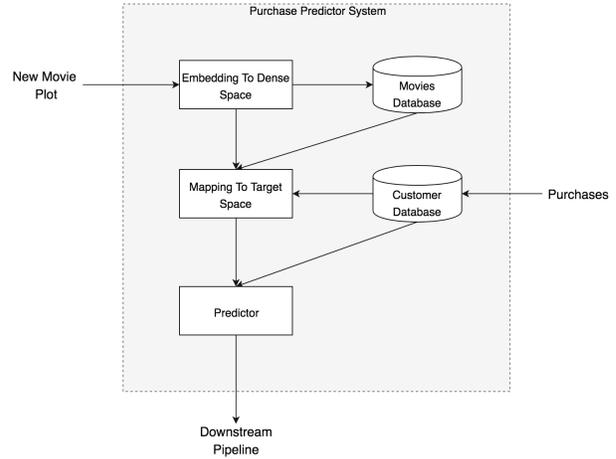

**Figure 2. Overview of the prediction system**

It would be useful, to make cold-start recommendations for products that cross categories, to have a model in which B&B and WW end up, after training, in the 'neighborhood' of TGS *before* the movie has opened. A proposed solution in the literature is to 'replace' the dot-product by a distance function that satisfies the triangle inequality and that can more easily propagate product-to-product underlying similarity based on customers' actions. Distance and similarity are central to many machine learning algorithms. Collaborative Metric Learning [10] is a recently developed approach that blends metric learning techniques and collaborative filtering ideas to learn product content and customer representations in a latent content space in which the triangle inequality holds. Unlike CF models based on dot product, metric learning models satisfy the triangle inequality and have been shown more suitable to uncover fine-grained relationships among customers' preferences for products that cross established categories (probably at the expense of accuracy for products that fit nicely into established categories.)

We present a variant of the CML approach to predict purchase probabilities for novel products based on product content and binary (yes/no) customer purchase data. Like CML, we use product distance in the loss function of the deep architecture to train the model. Unlike CML, which learns users and products positions in the space by minimizing the distance between users and the products they consume, our model learns the products positions by minimizing the distance for products that co-occur in customers' histories.

We use a Siamese deep network architecture to learn the distance [6]. Existing DLM models learn to quantify the distance between instances by training on datasets structured as pairs of instances, and on binary labels that specify whether the two instances belong to the same category ([3], [4]). Those models learn a nonlinear transformation that projects the instances to a target space in which the metric distance between positive pairs is smaller than the distance between negative pairs. In our case, the metric in the target space can be learned by training on each of the customers' purchase histories structured as pairs of products that co-occur (positive pairs) or fail to co-occur (negative pairs) depending on the customer. The system predicts that customers will purchase a new product when the new product is similar to other products that they have purchased in the past, or to products that are associated by purchase transitivity.

In our proposed system, we first map sparse natural language product descriptions to a dense space using embedding-like deep models [2]. We then train a deep model that uses customers' histories to map embedded product vectors to a target space. The trained model brings closer together, in the target space, products that frequently co-occur in customers' histories. Instead of learning latent customer vectors like in [10], we reconstruct customer vectors based on past product purchases and a time discount function [13]. We do this to account for the fact that customers 'drift' in the product space because of changes in preferences and we would miss this important mechanism if we employed latent customer vectors. In addition, we incorporate customer frequency and time elapsed since the last purchase (recency) to estimate the purchase probability and to rank products in customers' choice sets.

In the inference step, we use the learned metric to calculate the distance between a new product and each and every one of the customers in the data. We use the calculated distance to each customer, as well as frequency-recency customer data, to predict the purchase probability for each combination of customer and new product (see Figure 1). This information is used to generate lists of comparable titles, evaluate audience composition and overlap with competitive titles, and build customer segments for personalized acquisition campaigns.

This work contributes to the literature in recommendation systems because it shows the application of a system to products before the product is built and available to the customers. Specifically, the contributions of this work include:

- A description of a model deployed in a production system that we are using to feed the analysis pipeline and the customer acquisition pipeline.
- A framework that combines feed-forward neural networks for embedding and for distance learning to estimate a nonlinear transformation to a target space in which distance is a better proxy of purchase probability.

- A linear model that combines customer-product distances in the target space with frequency-recency customer data.
- The results of a test for a group of movies that were released outside of the training window.

## 2. SYSTEM OVERVIEW

The high-level architecture of the recommendation system for theatrical movie openings is shown in Figure 2. The initial step involves writing a long-form description of the plot of a movie that is going to be released in the future, as well as a description of the plot of other movies that are opening at or around the same time. In addition to the plots, this step also requires collecting meta data about those movies, like movie genre, casting, etc. Importantly, we do not include in the description of the movie, or in the metadata, information about the franchise the movie belongs to. We do this because we want to place as much weight as possible on the plot description and the cast. Plots make franchises, not the other way around. The plots, the meta data, and the customer data are the inputs to the model. The output of the model is a purchase probability score for each customer in the database. The customer purchase probabilities are then fed downstream to the rest of the analytics and activation pipelines. Once the movie has opened, the actual purchase data is added to the training set and the model is retrained.

The first stage computes the high dimensional embedding of the movie plot (we use movie synopses and long movie scripts) and movie meta data (cast, movie genre) in a dense space. After tokenizing the movie plot, we prepare the data using a step that we call movie plot compactification. To compactify a movie plot, we create a sequence of words that alternates movie metadata elements with elements from the tokenized original plot. We then append the compactified plot to the original plot and use that as the input to the embedding stage. The embedding model [2] exploits associations between contiguous words ('spaceship' and 'alien') and associations between movie metadata items and plot words ('Jennifer Lawrence' and 'survival') to learn a dense representation of each word and each meta data item as fixed-length vectors in a high dimensional space. The final dense representation of the movie plot and associated meta data is computed as the average of all the high-dimensional vectors for all the words and meta data elements for each movie. This first stage is trained on data for hundreds of movies.

The second stage, the mapping, learns the distance metric between the embedded feature vectors in the dense space. The purpose of this stage is to learn a function that maps embedded feature vectors into a target space such that the L2 norm between two vectors in the target space represents the probability that the two movies they represent co-occurred in customers' purchase histories.

The third stage, the predictor, is a linear model that learns the purchase probability for each customer, P(y|**x**), given the inputs **x** that include the L2 distance between a new movie (in the target space) and the customer history (in the target space), as well as frequency-recency customer data.

The separation of the system in three different stages could negatively impact system performance but was helpful during the build stage because it allowed us to have different teams working simultaneously on different parts of the system. To guide development changes and overall system integration we calculate precision, recall and ranking loss for a number of customer segments that differ in frequency of moviegoing, number of tickets per transaction, etc. We use this information to identify the optimal operation point for the system. More importantly, we test the results of the system in our media campaigns by 'turning the system on/off' and measuring changes in watch time, engagement with video, relevancy of the video, shares, likes and other online metrics. Finally, after the movie has been released, we analyze the performance of the system by comparing the predictions to the actual customers' purchases.

In the next section, we describe in detail the second and third stages.

## 3. DEEP METRIC LEARNING IN THE PRODUCT SPACE

### 3.1 Distance Metric Learning

The distance learning stage estimates a function that maps the embedded movie plot (e) into a representation of the movie in a target space (z):

$$z_{movie} = f(e_{movie})$$

The main idea is to find a function that brings closer together in the target space those movies that are more likely to co-occur in customers purchase histories. When the function is then applied to a new plot, the local L2 neighborhood of the new movie in the target space should be a good predictor of which other movies are going to be in the customers histories *of those customers that are more likely to see the new movie*. Because of this, the L2 distance in the target space between a new movie and the vector that represents a customer's purchase history is a predictor of the purchase probability.

To learn the mapping function, we use a Siamese framework with two identical feed-forward multilayer neural networks comprised of dense layers of rectified linear activation units (ReLUs). The input to the Siamese framework for each training instance is a pair of embedded vectors and a 'one/zero' label that indicates whether a customer purchased the two movies. A contrastive loss function like the one used in [6] is used to learn the metric. The function penalizes large distances for positive instances and low distances for negative instances. The training algorithm backpropagates the gradients from the output to different layers of the Siamese network using mini-batch stochastic optimization.

### 3.2 Logistic regression model

The logistic model estimates the purchase probability based on the distance between the customer and the movie in the target space, as well as frequency(F)-recency(R) customer data.

$$logit(P(Y=1|z,F,R)) = \alpha + \beta\phi(z_{customer}, z_{movie}) + \gamma(F,R) + \varepsilon$$

The distance $\phi$ between the customer and the movie is calculated as the Euclidean distance between the movie vector in the target space and the customer vector in the target space.

$$\phi(z_{customer}, z_{movie}) = \|z_{movie} - z_{customer}\|_{L2}$$

The customer vector is calculated as the average of the movie vectors for all the movies in the customer's history, weighted by some discount time factor

$$z_{customer(i)} = \frac{\sum_{n \in \Omega(i)} z_n e^{-\delta t(n)}}{\sum_{n \in \Omega(i)} e^{-\delta t(n)}}$$

## 4. SYSTEM IMPLEMENTATION

To prepare the training dataset for the Siamese network, we use the customer purchase data to build the Cartesian product of the purchase history of each customer, and the complete list of movies in the dataset.

$$\{movie_{customer(i)} \times all_{movies}\}$$

Each row in the dataset represents a possible combination of customer, movie purchased by that customer, and movie from the movie universe. We label the rows as 'one' if both movies were purchased by the customer. Otherwise, we label the rows as 'zero'. To avoid using the future to predict the past, we retain only the instances where the movie in the first column was released before the movie in the second column. To train the Siamese network, we append the fixed-length embedding vectors for each of the two movies in each row. We use the resulting dataset to train the Siamese network and learn the mapping function to the target space.

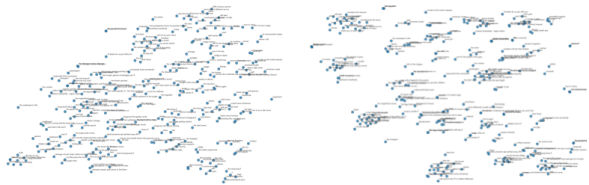

**Figure 3. Left, movie clusters based on truncated movie plots and movie meta data. Right, movie clusters based on complete movie plots and movie meta data**

To prepare the training dataset for the Logistic regression model, we calculate, for each customer, the vector representing the purchase history of that customer in the target space, discounted by some time factor. We then calculate, using the trained Siamese network, the distance in the target space between the vector that represents each of the movies in the dataset, and the customer vector. The customer vector is computed with the purchase history of that customer *prior to release of the movie*. We repeat for each combination of movie in the dataset and customer. We include additional columns to capture frequency and recency customer data. We label each row as 'one' if the movie was actually part of the customer history. Otherwise, we label the row as 'zero'.

The dataset used to train the logistic model shows a much larger proportion of negative cases than positive cases. This is to be expected since moviegoers on average only go to a number of movies a year. We optimize the hyperparameters of the logistic goal function in order to place different weight on false positives and false negative cases. We do this in order to find the right balance between precision, recall and movie ranking in a test hold out sample.

During the inference stage, we load in memory the weights of the embedding model, the weights of the mapping function learned by the Siamese network, and the weights of the logistic predictor. We use the embedding and the mapping function to calculate the distance in the target space between the customer vector, and the vector that represents the movie for which we want to make inferences. And we use the logistic predictor to estimate the purchase probability for each customer in the database. After doing this for a number of target movies (e.g., all the movies that open during a given week or month of the year), we are able to compute the predicted movie ranking (and associated probabilities) for each customer, as well as the characteristics of the inferred audience for each of the target movies (mix of casual vs high frequency audience, demographics, geographical distribution, etc).

## 5. RESULTS

We evaluate the performance of the model by analyzing the AUC of the model for movies in the training set, and for new movies not included in the training set (released in the following two months). We analyzed two cases. In the 'baseline' case, we only use customer recency and frequency, but not distance, to predict ticket purchase. This is the baseline model that would be available before the deep component is added to the model. In the 'deep architecture' case, we use recency, frequency *and* the Euclidean distance between the movie and the customer to predict ticket purchase. As can be seen in Table 1, the deep architecture case over performs not only the 'baseline' case in the training set movies (gain = +5 percent point) but also in the test set (gain = +4 percent point). This suggests that we can use the proposed architecture to learn the metric between vectors in a dense (embedded-like) space and that distance in this space is a predictor of purchase probability.

### 5.1 Visualizing the Impact of Movie Plots

Product descriptions improve the performance over the baseline model because of the ability of the deep component to generalize the learnings to movie plots that have never occurred in the past. To understand the significance of the impact of the deep component, we compared the calculated position in a two dimensional x-y chart for all the movies in the database. We compared two cases. In the first case, we used truncated movie plots. in the second case, we used the full movie plots. As seen in Figure 3, when we use the full movie plots we are able to reproduce distinct movie clusters.

### 5.2 Comparable-Movies System Results

A comparable-movies system is a system that predicts the audience composition of a movie in terms of other (past) movies. Being able to predict audience composition in terms of past movies is important for movie studios to architect successful franchises, produce successful movies, identify optimal release windows, and execute on-target marketing campaigns.

The problem of finding comparable movies using customer data can be solved using the recommendation system described in this paper. After the recommendation system has computed the purchase probabilities for a new movie, we can use those probabilities to select a segment of the customer base that is more likely to attend the movie, and compute the top movies that bubble up in customers' histories for that segment. After the movie is released in the theaters, we can compare the predicted list with the actual list. If the recommendation system is imprecise or inaccurate, one will find discrepancies between the two lists of comparable movies.

Table 2 and 3 show predicted and actual list of comparable movies for two titles: The Greatest Showman and Ferdinand. When applied to The Greatest Showman, the comparable-movies system identified eight out of the top ten movies that bubble up in customers' histories for those customers that went to The Greatest Showman. For Ferdinand, the system identified six out of ten. These initial results suggest that deep model architectures can be used to learn customer preferences based on content information and customers' purchase histories, and that movie studios can use deep model architectures to influence strategy decisions and execute on-target and personalized marketing campaigns.

**Table 1: AUC for the model in the training set and AUC for the model applied to new movies.**

|  | Training set AUC | New movies AUC |
|---|---|---|
| **Baseline** | 0.63 | 0.60 |
| **Deep Architecture** | 0.68 | 0.64 |

**Table 2: Comparable-Movies System Results for The Greatest Showman**

| PREDICTED | ACTUAL |
|---|---|
| Beauty & the Beast | Beauty & the Beast |
| La La Land | La La Land |
| Cinderella | Cinderella |
| Wonder Woman | Pitch Perfect 2 |
| Guardians of the Galaxy 2 | Hidden Figures |
| Pitch Perfect 2 | Annie |
| Passengers | Passengers |
| Fantastic Beasts and … | Wonder Woman |
| Annie | Fantastic Beasts and … |
| A Dog's Purpose | Wonder |

**Table 3: Comparable-Movies System Results for Ferdinand**

| PREDICTED | ACTUAL |
|---|---|
| The Boss Baby | The Boss Baby |
| The Secret Life of Pets | Sing |
| Sing | The Secret Life of Pets |
| Finding Dory | Moana |
| Moana | Coco |
| Zootopia | Cars 3 |
| Hotel Transylvania 2 | Trolls |
| Inside Out | Despicable Me 3 |
| Despicable Me 3 | Captain Underpants |
| Beauty & the Beast | Finding Dory |

## 6. CONCLUSION

Understanding customer preferences and shifts in underlying tastes is important for movie studios. Machine learning models developed to optimize customer experience for online movie streaming services might not be adequate for traditional studios because of data limitations. We presented a system that learns to measure distances in a product space to predict purchase probabilities. The system is based on a deep neural network architecture that trains on customer purchase histories and on dense representations of movie plots. Initial experiments show material gains over traditional frequency-recency models.

Movie studios routinely evaluate investments in movie scripts and talent. Deep models have traditionally existed inside the domain of 'big tech'. Democratization of deep analytic models (through, e.g., open source libraries like Tensorflow) has lowered the barriers of entry for traditional studios and main street brands in general. By building on the rich literature in recommendation systems, studios can not only use machine learning to improve the odds when evaluating movie scripts, but develop end-to-end machine learning pipelines to power effective acquisition and personalization marketing campaigns. Deep analytic models that blend collaborative filtering techniques and distance learning techniques show promise in cold-start situations for movies that cross 'genres' or movies that do not fit neatly into established categories. Models can be enriched with customer frequency and recency data. Current active work that explores how to embed frequency and recency into the metric itself (as opposed to overlaying frequency and recency through, e.g., a separate stage) should improve model accuracy and general performance in real world situations with a dense spectrum of customer tiers, each with their own preferences and trajectories.

**Acknowledgment**

We appreciate Andy Hsieh and Hao Wang for thoughtful comments and feedback, Xun Wang, Brandon Schelenker, Anthony Helguera and Christopher Pufall for invaluable support with data preparation, and Sarah Nooravi, Isaiah Yoo and Samuel Covert for support with movie script encoding.